\documentclass[rsi,twocolumn,aip]{revtex4}

\usepackage[dvips]{graphicx}

\begin{document}
\date{\today}
\title{A Compact Apparatus for Studies of Element and Phase-Resolved Ferromagnetic Resonance}
\author{D.A.~Arena\footnote{e-mail address: darena@bnl.gov}}
\affiliation{National Synchrotron Light Source, Brookhaven National Laboratory, Upton, NY, U.S.A.}

\author{Y.~Ding}
\affiliation{National Synchrotron Light Source, Brookhaven National Laboratory, Upton, NY, U.S.A.}

\author{E.~Vescovo}
\affiliation{National Synchrotron Light Source, Brookhaven National Laboratory, Upton, NY, U.S.A.}

\author{S.~Zohar}
\affiliation{Material Science Program, Department of Applied Physics, Columbia University, New York, NY, U.S.A.}

\author{Y.~Guan\footnote{present address: SoloPower Inc., 5981 Optical Court, San Jose, CA 95138}}
\affiliation{Material Science Program, Department of Applied Physics, Columbia University, New York, NY, U.S.A.}

\author{W.E.~Bailey}
\affiliation{Material Science Program, Department of Applied Physics, Columbia University, New York, NY, U.S.A.}

\begin{abstract}
We present a compact sample holder equipped with electromagnets and high frequency transmission lines; the sample holder is intended for combined x-ray magnetic circular dichroism (XMCD) and ferromagnetic resonance measurements (FMR).   Time-resolved measurements of resonant x-ray detected FMR during forced precession are enabled by use of a \emph{rf} excitation that is phase-locked to the storage ring bunch clock. Several applications of the combined XMCD + FMR technique are presented, demonstrating the flexibility of the experimental design.
\end{abstract}

 \maketitle

\section{Introduction}
Among the most powerful methods developed to examine magnetic materials are ferromagnetic resonance (FMR) and x-ray magnetic circular dichroism (XMCD).  FMR is certainly the more venerable technique and has been used to examine numerous phenomena (\emph{e.g.} gyromagnetic ratios, magnetic anisotropy energies, damping and relaxation processes \cite{Farle:1998ts,Lindner:2003la,Simanek:2003fy}, \emph{etc.}).  In thin films, these capabilities have been extended to investigate issues such as magnetic reorientation transitions, substrate effects in magneto-crystalline anisotropy, and coupling in multilayer systems.  FMR is a well-established approach and is supported by a firm theoretical foundation that can be applied in the interpretation of experimental spectra.  

By contrast, XMCD is a much more recently-developed technique \cite{Erskine:1975zh,Chen:1990xh}.  XMCD can also be applied to examine a wide class of issues including overlapping topics such as magnetic anisotropies and spin-reorientation transitions.  Moreover, because XMCD is based on core-level spectroscopy, it provides elemental sensitivity and can be used, with appropriate care and via the application of sum-rules analyses, to determine absolute spin ($\mathbf{\mu_{s}}$) and orbital ($\mathbf{\mu_{l}}$) moments \cite{Thole:1992pr,Carra:1993bs,Chen:1995dp}.  These characteristics make XMCD an indispensable tool for the investigation of complex ferromagnetic systems such as alloys and other compounds and layered thin-film structures.  

The combination of these two techniques has been a goal of several research groups over the past few years and a few different implementations of XMCD + FMR have been developed.  The approaches used in these efforts typically fall into two groups: static, or time-averaged, measurements and time-resolved techniques.  The time-averaged methods generally operate with continuous wave (\emph{cw}) microwave excitations.  While the details vary, in a typical implementation a microwave field is used to excite uniform precession modes, and the change in the projection of an elemental moment ($\mathbf{\Delta\mu_{Z}}$) parallel to the incident x-ray wave vector ($\mathbf{k_{i}}$) is monitored as either the microwave frequency or the magnetic field is swept through resonance \cite{Boero:2005hs,Goulon:2005zv}.  The absorption of the \emph{cw} excitation results in a change in $\mathbf{\Delta\mu_{Z}}$.  In general, in such time-averaged measurements the \emph{rf} excitation bears no particular phase relation with the arrival of the x-rays and hence all frequencies are accessible.  

Time-resolved XMCD with continuous wave excitations, have been implemented to image sub-GHz gyrotropic motion of vortex dynamics in mangetic nanostructures \cite{Puzic:2005jl}.  The time-resolved approach has been extended by our group \cite{Bailey:2004tw,Arena:2006kc,Guan:2006ul,Guan:2007jw,Arena:2007qh}and other researchers \cite{Martin:2008ud} to examine GHz-range FMR with elemental-specificity.  An advantage of time-resolved techniques is that the phase information between the \emph{cw} excitation and the precessing magnetic moments in the sample is preserved, permitting a flexible exploration of the multi-dimensional phase space (excitation frequency, phase of the response, photon energy, and applied field) inherent in the experiment.  

We present in this article an experimental apparatus that allows for the combination of time-resolved XMCD and FMR. We designed a flexible measurement system, which is suitable for x-ray absorption spectroscopy (XAS), as well as conventional XMCD and \emph{in-situ} FMR.  Time-resolved studies are enabled by the use of phase-locked \emph{rf} generation electronics.  The hardware is compact as it fits in the bore of a standard 4.5'' UHV flange.  The measurements presented in this article were undertaken at the soft x-ray beam line 4-ID-C at the Advanced Photon Source (APS) at Argonne National Lab.  The beam line is equipped with an innovative hybrid electromagnetic circularly polarizing undulator (CPU) \cite{Freeland:2002cy}, which can provide arbitrary x-ray polarization (linear horizontal or vertical, right or left circular). However, we stress that the apparatus and electronics described in this article are compact and easily portable to other beam lines and synchrotron facilities.

\section{Experimental Setup}
\subsection{Hardware: Sample Holder and Vector Magnet}
FMR in the linear regime implies small angular motions.  Therefore, regardless of the geometry of the measurement (\emph{e.g.} transverse or longitudinal), the requirements on sample and beam stability are severe.  Furthermore, near resonance, the phase of the magnetization motion changes rapidly, and thus the field homogeneity and stability over the area sampled by the photon beam is critical. To address these restrictions, we designed a custom sample holder to house the sample, \emph{rf} transmission lines and waveguides, and an in-plane vector magnet which provides vertical and horizontal fields.  In addition to these restrictions, a goal of the design was compactness as the entire assembly is intended to fit into the bore of a \emph{x,y,-z} translation stage with a clear opening of 2.5".

\begin{figure}[!]
  \begin{center}
  \includegraphics[height=25em]{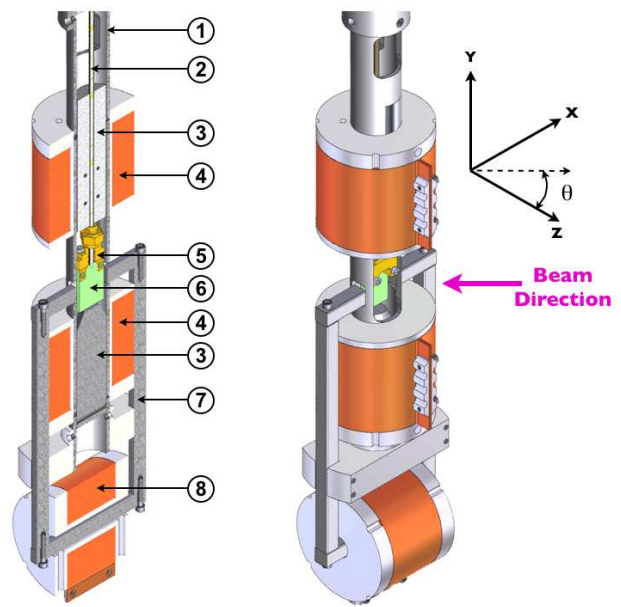}
  \caption{(Color Online) View of vector magnet and sample environment.  Right Panel: perspective with coordinate axes and indicating beam direction. Left panel: Cross section through the sample plane: 1: support tube; 2: high frequency coaxial cable with SMA termination; 3: top and bottom vertical pole pieces; 4: top and bottom electromagnet coils; 5: end launcher; 6: sample; 7: yoke for horizontal magnetic field; 8: electromagnet coil for horizontal field.}
  \label{MagnetFig}
  \end{center}
\end{figure}

Fig. ~\ref{MagnetFig} presents a perspective view of the custom sample holder as well as a cross section through the plane containing the magnets and the sample.  The coordinate system is chosen so the sample lies in the \emph{x-y} plane, with the sample normal along the \emph{z} direction, while the incident synchrotron radiation is restricted to the \emph{x-z} plane.  The angle of the photon beam with respect to the sample normal is varied by rotating the sample manipulator about the \emph{y}-axis. 

RF fields were delivered to the sample using a SMA end launch adapter to a grounded coplanar waveguide (CPW) transition, similar to those used in pulsed inductive microwave magnetometry\cite{Kos:2002rw,Reidy:2003rc}.  The orientation of the waveguide's center conductor is along the \emph{y}-axis and thus the \emph{rf} field ($\mathbf{H_{\emph{rf}}}$) is in-plane and along the \emph{x}-direction. The end launch adapter (Southwest Microwave 292-06A-5 or similar), 1/2" in width, is screw-mounted to through-holes on the CPW, and has a coaxial center pin diameter and dielectric diameter size-matched to the CPW center conductor and ground shield to ground shield spacing, respectively.   The coplanar waveguide was formed from 10 mil thick, double-sided Au plated Rogers laminate (R5880), with $\epsilon_{r}=\textrm{2.2}$, center conductor width 21.4 mil, and ground-shield to ground shield spacing 36.5 mil, with dielectric spacing on each side of the center trace of 7.5 mil.  The total length of the center trace is only 327 mil. 

An innovative feature for the transmission experiment was the formation of a CPW transparent to x-rays, allowing x-rays to pass through the magnetic film and be measured at a photodiode behind the sample.  We have accomplished this by drilling a 10 mil ($\simeq\textrm{250}\mu m$) diameter pinhole through the center trace 50 mils above the lower edge of the CPW, leaving $\sim$ 13 mils of center trace on either side of the pinhole.  Thin film samples, deposited on Si$_{3}$N$_{4}$ membranes, are mounted, using epoxy, over the hole.  The input \emph{rf} power is reflected by a dead short at the end of the CPW, formed by terminating the dielectric spacing, or (equivalently) allowing the Au ground shield and center conductor to continue for a 10 mil width at the CPW edge opposite the end launch.  This places the optically accessible region of the sample at a voltage node and \emph{rf} field maximum.  An upper limit for the \emph{rf} field can be given by the DC field from a planar conductor,

\begin{equation}
B^{rms}_{rf}={\mu_{0}\over 2 W}{\sqrt{P\over Z_{0}}}
\end{equation}

where $W=\textrm{563}\mu m$ is the CPW center conductor width, and $Z_{0}\simeq\textrm{50}\Omega$ is the characteristic impedance of the waveguide, where the expression assumes film spacing from the CPW much less than the CPW width, reasonably well justified in the present case.  For the maximum input power of 30 dBm (1 W), and present center conductor width, this is $B^{rms}_{rf}\leq\textrm{163.6 }\mu \textrm{T}$, or $H^{pp}_{rf}\leq \textrm{2.3 Oe}$.  In the vicinity of the hole, the \emph{rf} field will likely be less, although more detailed simulation would be necessary to estimate its value.

The sample environment contains two electromagnets oriented along the \emph{y}-direction.  A high bandwidth coaxial \emph{rf} cable passes axially through a hole in the top pole piece and up to a \emph{rf} vacuum feedthrough (not shown).  These two electromagnets provide a magnetic field referred to as the vertical bias field ($\mathbf{H_{B}}$), which is normal to both to $\mathbf{H_{\emph{rf}}}$ and the incident photon beam.  $\mathbf{H_{\emph{rf}}}$ forces the magnetization in the sample to precess about $\mathbf{H_{B}}$ (or, more accurately, about $\mathbf{H_{eff}}$, which is the sum of $\mathbf{H_{B}}$ and the anisotropy and dipolar fields).  At a fixed driving frequency, the precession of $\mathbf{M}$ can be tuned by varying $\mathbf{H_{B}}$.  A third electromagnet is connected to a \emph{C}-shaped yoke and provides an in-plane magnetic field along the \emph{x}-direction (\emph{i.e.} horizontal and orthogonal to $\mathbf{H_{B}}$).  This field ($\mathbf{H_{x}}$)is used to acquire conventional XMCD spectra and element-specific hysteresis curves.  

The pole pieces for all three electromagnets were machined from conventional low carbon-content steel.  To reduce the internal stress caused by the facbrication processes, the pole pieces were annealed in an inert atmosphere at 960$^{\circ}$ C for 8 hours, followed by a slow cool down.  This procedure produced relatively ``soft" magnetic cores with low remanent magnetization.  In the linear region of the magnetization curve, the slope of the $\mathbf{H}$ \emph{vs.} applied current curve is 32 Oe / amp and 38 Oe / amp for the vertical and horizontal electromagnets, respectively.  Ohmic heating of the coils, combined with the lack of conductive and convective cooling in vacuum, limits the sustainable current to $\sim$3 amps.  However, the coils are equiped with cooling strips made of thin Cu foil.  Future modifications will add \emph{in-situ} cooling to the three electromagnets, as well as sample cooling for temperature-dependent measurements.

A significant advantage of the experimental system presented is the simplicity of sample preparation.  The samples are planar ferromagnetic thin films, deposited onto commercially available, 100 nm thick Si$_{3}$N$_{4}$ membranes.  In the spectral range of interest (\emph{L} edges of 2$^{nd}$ row transition metals and \emph{M} edges of rare earth elements) such membranes have a high degree of transmission ($\gtrsim$ 75\%).  The membranes are supported by Si frames (500 $\mu$m thickness) and these are attached to the waveguide using an insulating adhesive.  

Spectroscopic x-ray absorption data and timing scans are acquired in transmission mode.  Assuming an appropriate selection of materials and sample thickness (required to avoid saturation effects), transmission measurements have a significant advantage over electron or fluorescence yield techniques in improved signal-to-noise ratio.  A related benefit is simplified detection utilizing a standard soft x-ray photodiode operated in current amplification mode.  Also, in contrast to pulse counting, noise reduction via lock-in amplification is implemented in a straightforward fashion.

\subsection{Microwave Excitation and Timing Electronics}
\begin{figure}[t]
 \begin{center}
  \includegraphics[height=25em]{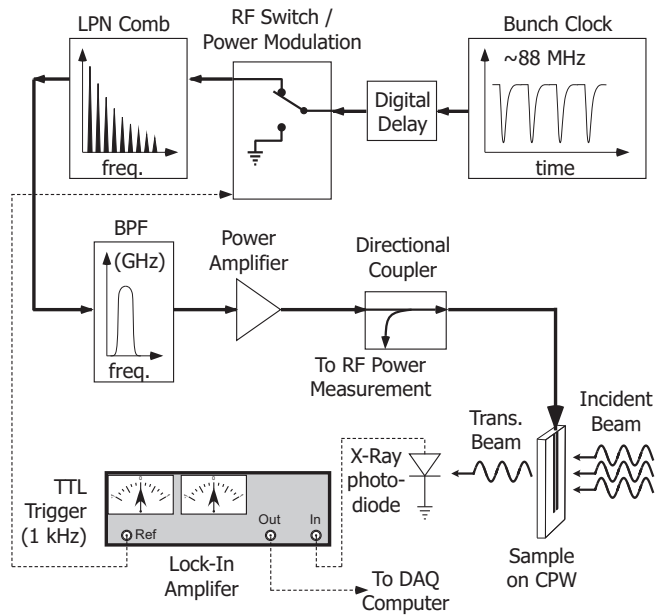}
  \caption{Simplified schematics of \emph{rf} and signal detection electronics.  The generation of the phase-locked \emph{rf} excitation follows heavy solid lines while the detection of the x-ray signal proceeds along the light dashed lines.  Note that the lock-in detection branch may be by passed by setting the TTL trigger to ``high.''}
  \label{rfSchematic}
  \end{center}
\end{figure}

In our measurements, as $\mathbf{H_{\emph{rf}}}$ drives the precession of $\mathbf{M}$, the projection of $\mathbf{M}$ along the photon beam direction ($\mathbf{M_{proj}(t)}$) is sampled stroboscopically by the x-ray pulses.  This requires an \emph{rf} signal in the microwave region of the spectrum (in our case, $\sim$1 to 4.5 GHz) that maintains a well-defined phase relationship with the photon bunch clock.  Such phase-locking imposes a further restriction that the \emph{rf} excitation frequency must be a harmonic of the bunch clock frequency. Fig. ~\ref{rfSchematic} presents a simplified schematic of the electronics developed to generate this \emph{rf} signal.  The path of the microwave generation is indicated by the heavy solid line while the x-ray photodiode and lock-in output and control signals are depicted with a light dashed line.  For clarity, intermediate attenuation and amplification conditioning stages have been ommitted.


The \emph{rf} generation sequence starts with the photon bunch clock.  For our time-resolved measurements, we utilize special operating mode 4 (SOM-4) at the APS, which is available several times a year.  In SOM-4, every fourth \emph{rf} bucket in the storage ring is populated with electrons, which results in an x-ray pulse repetition frequency of $\sim$88 MHz.  The bunch clock signal is fed into a local digital delay generator.  After the delay stage, a high-bandwidth switch, triggered by a periodic TTL signal that also serves as the reference for the lock-in amplifier of the detection circuit (see below), modulates the signal on and off.  An amplified version of the delayed bunch clock signal is directed into a low phase noise comb generator (LPN comb), which generates the harmonic spectrum of the 88 MHz input signal.  A tunable narrow band pass filter (BPF) is used to select the desired harmonic of the input signal. After the final amplification stage in the circuit, a directional coupler is directs the signal to the CPW.  The reflected signal is detected with a microwave diode or \emph{rf} power meter at the auxiliary port of the coupler.  

For fixed-frequency, conventional FMR measurements, the reflected \emph{rf} power is monitored at the directional coupler while sweeping $\mathbf{H_{B}}$.  For the XMCD measurements, the signal from the photodiode is amplified by use of a current preamplifier (not shown).  To improve the signal to noise ratio in the small-amplitude XMCD+FMR measurements, the power of the \emph{cw} excitation is modulated at $\sim$1 kHz by the \emph{rf} switch, located between the bunch clock and the LPN comb.  The signal from the current pre-amplifier is directed to a lock-in amplifier to improve the signal to noise ratio.  Note that the \emph{rf} power modulation and noise suppression via lock-in amplification can be bypassed by setting the the TTL control signal to the \emph{rf} switch to``high."  In this configuration, the signal level from the photodiode during driven precession can be compared readily to hysteresis measurements, providing a simple angular calibration for signal levels.  

\section{Applications}
\subsection{Conventional XMCD Spectroscopy and Element-Specific Magnetometry}
\begin{figure}[!]
  \begin{center}
  \includegraphics[height=33em]{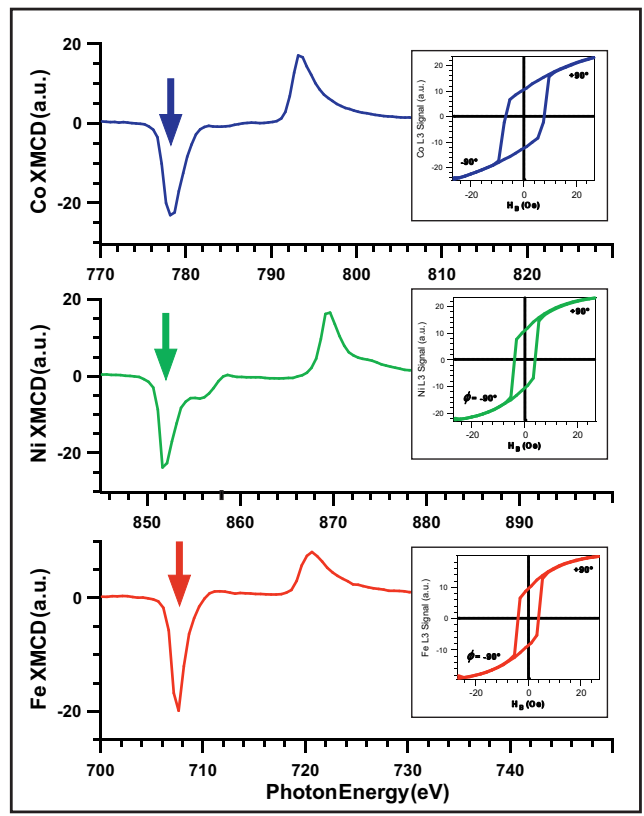}
  \caption{(Color online) XMCD spectra and element-speciÞc hysteresis measurements of the Ni$_{81}$Fe$_{19}$/Cu/Co$_{93}$Zr$_{7}$ trilayer. The hysteresis curves were measured at the photon energies indicated by the arrows (707.5eV for Fe, 851.5 eV for Ni and 778 eV for Co). The XMCD signal levels at saturation in the hysteresis curves provide an angular calibration for time-resolved XMCD measurements. Reproduced from ref. \cite{Arena:2006kc}}
  \label{staticXMCD}
  \end{center}
\end{figure}

As mentioned, the design of the apparatus depicted in Fig. 1 is quite flexible.  By using the horizontal electro-manget (item 8 in Fig. \ref{MagnetFig}), $\mathbf{H_{x}}$ can be used to provide an alternating saturating field for soft ferromagnetic materials, permitting measurement of standard XMCD spectra in field-switching mode with fixed x-ray helicity.  An example of such spectroscopy is presented in Fig. \ref{staticXMCD}, reproduced from ref. \cite{Arena:2006kc}.  The figure shows the XMCD spectra, acquired in transmission mode, of a magnetic trilayer [(Si$_{3}$N$_{4}$ substrate/Ni$_{81}$Fe$_{19}$ (25 nm)/Cu(20 nm)/Co$_{93}$Zr$_{7}$ (25 nm)/Cu cap(5 nm)], similar to technologically important giant magneto-resitance (GMR) stacks found in many modern magnetic field sensors.  The spectra were acquired in a single sweep with a reasonable dwell time per point ($\sim$1 s). High-quality XMCD spectra, clearly showing the spin-orbit split  $L_{3} ~(2p_{3/2})$ and  $L_{2} ~(2p_{1/2})$ core levels, are acquired from all three magnetic elements in the multilayer structure.   The low noise in the data indicates that the vibration of the sample holder relative to the photon beam and the photodiode is low.

The insets to Fig.~\ref{staticXMCD} show element-specific hysteresis curves acquired by settting the photon energy to the $L_{3}$ absorption edges of Fe, Ni or Co and then sweeping $\mathbf{H_{x}}$.  One important benefit of employing a core-level spectroscopic technique is immediately apparent in comparing the hysteresis loops of the Fe and Ni edges with the loop measured at Co $L_{3}$ edge: the Co$_{93}$Zr$_{7}$ clearly has a larger coercive field than the Ni$_{81}$Fe$_{19}$ layer.  More relevant to the element-resolved FMR scans (presented below) is the angular calibration provided by the hysteresis measurements; the total magnitude of the photodiode signal in sweeping the field from $\pm \mathbf{H_{x,max}}$ corresponds to a change of the direction magnetization by 180$^\circ$ (assuming the magnetization of the sample is saturated at the extremal values of the applied field).

\subsection {\emph{In-Situ} FMR Spectroscopy}
The electronics and hardware described above can be used in a straightforward fashion to measure conventional FMR spectra.  In these experiments, a sinusoidal \emph{rf} signal in the frequency range of $\sim$1-4.5 GHz is introduced along the high bandwidth coaxial transmission line. The reflected \emph{rf} power is monitored as $\mathbf{H_{B}}$ is varied.  Measurements are typically conducted using a lock-in amplifier to improve the signal-to-noise ratio.  With this set-up, the resonance fields and FMR linewidths can be determined before proceeding with the time-resolved measurements.  Examples of such \emph{in-situ} measurements can be found in references \cite{Arena:2006kc,Guan:2007jw}.  The conventional FMR spectra presented in these references are static measurements which integrate the absorbed power throughout the sample.  Additional details on the motion of elemental moments is revealed by utilizing XMCD and time-resolution. 

\subsection{Measurement of Precession Orbits}
\begin{figure}[t]
  \begin{center}
  \includegraphics[width=0.5\textwidth]{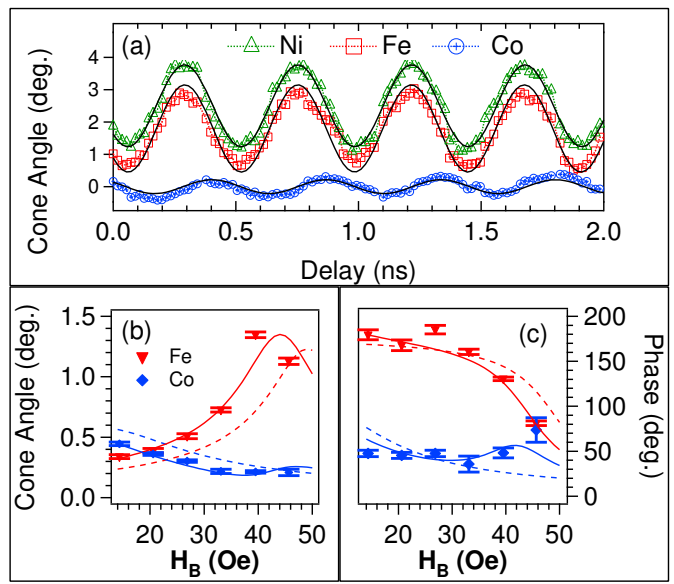}
  \caption{(Color online) Panel (a): representative time-resolved XMCD data of a Ni$_{81}$Fe$_{19}$/Cu/Co$_{93}$Zr$_{7}$ trilayer undergoing forced precession at 2.3 GHz.  The solid lines are fits to a sinusoidal function.  Panel (b): extracted values for the amplitude of the forced precession orbits, resolved to the Fe in the Ni$_{81}$Fe$_{19}$ layer or the Co in the Co$_{93}$Zr$_{7}$ layer.  Panel (c): the corresponding phase of the precession orbits.  In panels (b) and (c), the dashed lines are simulations based on independent layers while the solid lines are simulations which assume weak coupling.  See reference \cite{Arena:2006kc} for details.}
  \label{TriLayerPrecession}
  \end{center}
\end{figure}

With use of an \emph{rf} excitation that is phase-locked with the photon bunch clock at the synchrotron, the projection of the precessing elemental moments can be sampled stroboscopically at a fixed delay.  The time-varying projection of the precession orbit is recorded by sweeping the variable delay of the \emph{rf} relative to the photon bunch clock.  An example is presented in Fig.~\ref{TriLayerPrecession} (adapted from ref. \cite{Arena:2006kc}).  The  data were acquired from the same Ni$_{81}$Fe$_{19}$/Cu/Co$_{93}$Zr$_{7}$ trilayer sample, at $\mathbf{H_{B}}$ = 40 Oe, near the main resonance for the Ni$_{81}$Fe$_{19}$ layer at the excitation frequency of 2.3 GHz.  

The variation in the signal measured in the element-specific hysteresis loops provides an angular calibration for the oscillatory signal.  As the functional form of the precession orbit, projected onto the photon beam direction, is a simple sinusoid, a fit to the data results in high-precision determination for the amplitude and phase of the orbit.  The amplitude of the oscillation has been measured from $\sim$1.5$^\circ$ down to $\sim$0.1$^\circ$, with an estimated error as low as $\sim$0.1$^\circ$.  The estimated error on the phase of the oscillation (relative to the photon bunch clock) can be as low as 2$^\circ$ of the full oscillation period, which corresponds to an overall time resolution of 2 ps at the 2.3 GHz excitation frequency used in this example.  This is considerably smaller than the $\sim$70 ps bunch length (FWHM) of the photon bunches from the storage ring.

\subsection{Phase-Resolved XMCD Spectroscopy}
\begin{figure}[b]
  \begin{center}
  \includegraphics[width=0.3\textwidth]{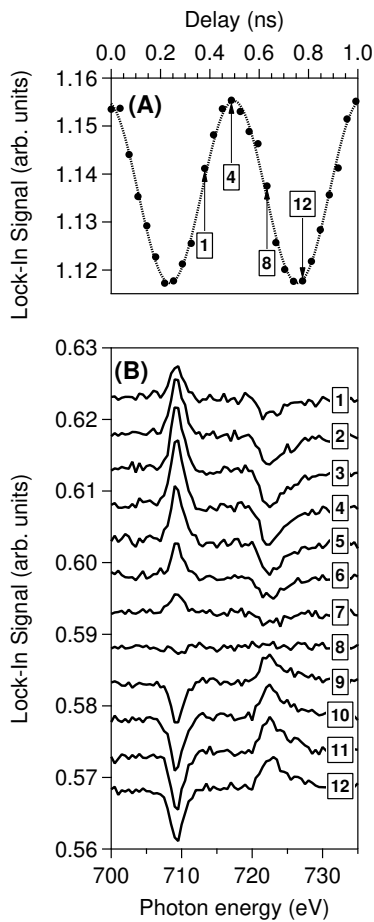}
  \caption{Panel (A): Timing delay scan acquired at 707 eV (Fe \emph{L$_3$} edge) from a Ni$_{81}$Fe$_{19}$ (25 nm) sample at an excitation frequency of 2 GHz. Panel (B): Photon energy scans at the delay points indicated in the top panel.  Note the variation in intensity of the dichroic signal, as well as the reversal of the dichroism on the \emph{L$_3$} and  \emph{L$_2$} edges.}
  \label{TR_spectros}
  \end{center}
\end{figure}

By selecting a specific time delay between the \emph{rf} excitation and the photon bunch clock,  the relative phase of the oscillations, and hence $\mathbf{M_{proj}(t)}$, can be held constant, permitting variation of other external parameters.  Fig. ~\ref{TR_spectros} presents one option, where the photon energy is swept through relevant absorption edges of the sample while $\mathbf{H_{B}}$ is held constant.  In this example, the sample consists of a single ferromagnetic layer, Ni$_{81}$Fe$_{19}$ [25 nm], with a 5 nm Cu cap.  Panel (A) of Fig. ~\ref{TR_spectros} (top) shows a representative timing scan acquired at the Fe $L_{3}$ edge near the resonance condition for the Ni$_{81}$Fe$_{19}$ film at \emph{f}=2 GHz.  The arrows in the figure point to specific delay values for the energy scans presented in panel (B).  The points are selected to span more than half a cycle of the precession orbit, and thus are assured to cover positive and negative extrema (near points 4 and 12, respectively) and an antinode (near point 8).

Panel (B) in Fig. ~\ref{TR_spectros} (bottom) presents the energy scans across the Fe $L_{3}$ and $L_{2}$ edges.  The spectra were acquired at fixed helicity of the CPU and with a point spacing of 0.5 eV and in a single scan of $\sim$10 minutes.    The data clearly show the change in the sign of the XMCD signal between the $L_{3}$ and $L_{2}$ edges, and initially the dichroism of the $L_{3}$ edge is positive.  The overall intensity of the dichroism starts off rather low, and increases in intensity as the delay value approaches point 4, close to the positive maximum of the oscillating $\mathbf{M_{proj}(t)}$.  After passing through this maximum, the intensity of the XMCD signal decreases and disappears at around point 8, where $\mathbf{M_{proj}(t)} = 0$.  Upon increasing the delay delay between the \emph{rf} and the photon bunch clock, the dichroism signal remerges, but now, as  $\mathbf{M_{proj}(t)} < 0$, the sign of the dichroism is reversed.  With a further increase of the delay, the intensity again increases, and the magnitude reaches another maximum at the negative extrema, near point 12 in Panel (A) of Fig ~\ref{TR_spectros}.

The data presented in Fig. ~\ref{TR_spectros} clearly show that phase-resolved XMCD spectra can be acquired efficiently.  With no significant changes to the apparatus, high-quality XMCD spectra can be acquired at specific points of the precession orbit.  In the future, by use of sum-rules analyses on these spectra, the spin-orbit ratio of the precessing elemental moments can be monitored with the same level of phase precision ($\sim$2$^\circ$) as has been demonstrated in the data in Fig.~\ref{TriLayerPrecession}.

\subsection{Element-Resolved Complex Susceptibility ($\chi$)}
\begin{figure}[t]
 \begin{center}
  \includegraphics[width=0.3\textwidth]{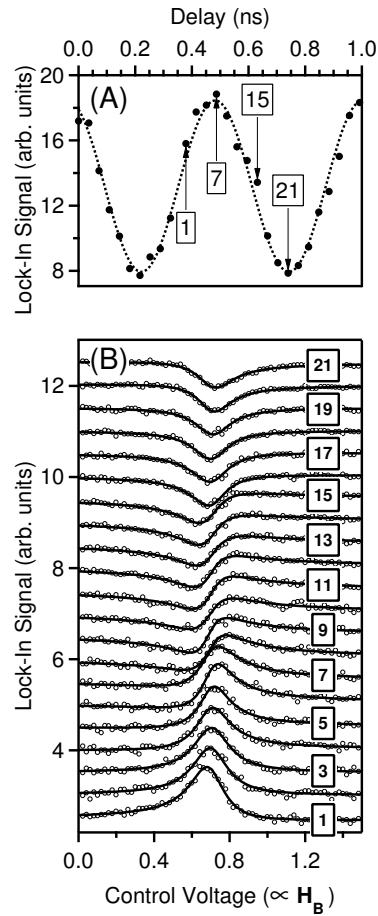}
  \caption{Panel (A): Timing delay scan acquired at 707 eV (Fe \emph{L$_3$} edge) from a a Ni$_{81}$Fe$_{19}$ (25 nm) sample. Panel (B): Magnetic field sweep scans at varying timing delays, corresponding to the point labeled in panel (A), with the photon energy fixed at the Fe \emph{L$_3$} edge.}
  \label{TR_Suscept}
   \end{center}
\end{figure}

Panel (A) of Fig.~\ref{TR_Suscept} presents the same type of delay scan as is found in Panel (A) of Fig.~\ref{TR_spectros}.  However, panel (B) of Fig.~\ref{TR_Suscept} presents a different variation of the experimental parameters.  Instead of keeping the relative phase and $\mathbf{H_{B}}$ fixed and varying the photon energy, in Fig.~\ref{TR_Suscept} the photon energy is held constant at the Fe $L_{3}$ edge while $\mathbf{H_{B}}$ is swept from 0 through the resonance condition.  The $\mathbf{H_{B}}$ scans are acquired at specific delay points, indicated by the arrows in panel (A) of  Fig.~\ref{TR_Suscept} .  As the delay values define a specific relative phase between the \emph{rf} and the photon bunch clock, the $\mathbf{H_{B}}$ scans can be interpreted as the phase and element-resolved complex magnetic susceptibility ($\mathbf{\chi=\chi' + i \chi''}$), in this case $\mathbf{\chi (Fe)}$ in the Ni$_{81}$Fe$_{19}$ film.  By varying the phase, the in-phase component of the susceptibility ($\mathbf{\chi '}$) or the out-of-phase component ($\mathbf{\chi ''}$) can be measured in an element-specific fashion.

As was the case with the energy scans, the $\mathbf{H_{B}}$ scans show a dramatic variation with the relative phase between the \emph{rf} and the photon bunch clock.  Panel (B) in Fig.~\ref{TR_Suscept} presents the data (open circles) and also a fit to a complex Lorentzian function (thin solid line).  At point 1, $\mathbf{\chi (Fe)}$ is asymmetric, with both real and imaginary contributions to the susceptibility.  At approximately point 3, the response to sweeping $\mathbf{H_{B}}$ is a symmetric and purely imaginary Lorentzian; at this condition, the response of the system is absorptive and 90$^\circ$ out-of-phase with the \emph{rf} driving field.  By point 12, at the nodal point of $\mathbf{M_{proj}(t)}$, the response is purely real and in-phase the \emph{rf}.  Beyond this nodal point, the response is again asymmetric until around point 19, where $\mathbf{\chi (Fe)}$ is again symmetric, although the reversal of the sign of $\mathbf{M_{proj}(t)}$ causes the Lorentzian function to be inverted.

\vspace{0.2 in}
\section{Outlook and Conclusions}
As can be seen in the various applications presented above, the combination of a compact vector magnet combined with an \emph{rf} excitation source that is phase-locked to the photon bunch clock produces a powerful tool to examine issues the dynamic response of scientifically- and technologically-relevant systems and materials. Static measurements of materials properties permit the identification of elements and even chemical species in a sample while the element-specific magnetometry can be used to establish magnetic anisotropies.  The resonance fields and linewidth, averaged across the sample, are easily measured via \emph{in-situ} FMR scans.  Extremely high-precision assessments of the precession orbit amplitude and phase are determined by timing delay scans at selected values of $\mathbf{H_B}$.  Finally, we have recently upgraded these capabilities by use of \emph{rf} power modulation and lock-in amplification.  These advances have in turn permitted implementation of time-resolved XMCD spectroscopy and measurements of element-resolved complex susceptibility ($\chi$).  

There are numerous issues that can be investigated by use of the experimental apparatus as presented (\emph{e.g.} origins of intrinsic\cite{Kambersky:2007ad,Gilmore:2007bf} and impurity damping\cite{Bailey:2001qf}, interfacial effects in multilayer structures\cite{Tserkovnyak:2002qe}, spin-transfer and spin-pumping effects\cite{Brataas:2004mb,Woltersdorf:2005yo}, \emph{etc.}).  These capabilities can be extended with relatively minor modifications, such as incorporation of sample cooling and implementation of a full return path for the \emph{rf} excitation.  These enhancements will facilitate investigations of an extended set of issues such as the origins of viscous damping in ferromagnets.

The use of synchrotron radiation for these measurements is critical, of course.  However, the properties of many 3$^{rd}$ generation storage ring sources sets an upper boundary on the frequency range accessible in these measurements.  The photon bunch length must be less than 1/2 of the oscillation period\cite{Horowitz:1989rz}.  For our measurements, the photon bunch length is $\sim$70 ps (FWHM), which sets an upper boundary of about 7 GHz.  For many systems, it would be desirable to extend this frequency range upwards.  Fortunately, shorter pulse lengths are available in special operating modes at some storage rings.  For example, the momentum compaction (or low $\alpha$) mode available at certain synchrotrons \cite{Feikes:2004qr} may permit measurements in the tens of GHz range.  

Support and Acknowledgements:  The technical assistance of Gary Nintzel and Tony Lenhard is gratefully acknowledged, as is the beam line support provided by David Keavney and colleagues at the Advanced Photon Source.  This work was partially supported by the Army Research Office with Grant No. ARO-43986-MS-YIP and the National Science Foundation with Grant No. NSF-DMR-02-39724. Use of the Advanced Photon Source was supported by the U.S. Department of Energy Office of Science, Office of Basic Energy Sciences, under Contract No. W-31-109-Eng-38.   The support of the NSLS under DOE contract No. DE-AC02-98CH10886 also is gratefully acknowledged.


\end{document}